\documentclass[a4paper,10pt]{article}
\usepackage{amsmath}
\usepackage{amssymb}
\usepackage{amsfonts}
\usepackage{graphicx}

\renewcommand{\vec}[1]{\mathbf{#1}}

\title{Stochastic Modeling in Systems Biology}
\author{Jinzhi Lei\thanks{Email: jzlei@mail.tsinghua.edu.cn}
\\
{}\\
{\small Zhou Pei-Yuan Center for Applied Mathematics}\\
{\small Tsinghua University, Beijing, 100084, China}}

\begin{document}

\maketitle

\begin{abstract}
Many cellular behaviors are regulated by gene regulation networks, kinetics of which is one of the main subjects in the study of systems biology. Because of the low number molecules in these reacting systems, stochastic effects are significant. In recent years, stochasticity in modeling the kinetics of gene regulation networks have been drawing the attention of many researchers. This paper is a self contained review trying to provide an overview of stochastic modeling. I will introduce the derivation of the main equations in modeling the biochemical systems with intrinsic noise (chemical master equation, Fokker-Plan equation, reaction rate equation, chemical Langevin equation), and will discuss the relations between these formulations. The mathematical formulations for systems with fluctuations in kinetic parameters are also discussed. Finally, I will introduce the exact stochastic simulation algorithm and the approximate explicit tau-leaping method for making numerical simulations. 
\end{abstract}

\section{Introduction}

Systems biology is an interdisciplinary science of discovering, modeling, understanding and ultimately engineering at the molecular level the dynamic relationships between the biological molecules that define living organisms\footnote{According to Dr. Leroy Hood, the first president of the Institute of Systems Biology, Seattle. http://www.systemsbiology.org/Systems\_Biology\_in\_Depth}. This field is increasingly hot in recent decades as modeling molecular systems is not only fascinating but also possible in the post-genomics science. At the molecular level, many cellular behaviors are regulated by genetic regulation networks in which the stochasticity is significant. To describe these stochastic chemical kinetics, stochastic modeling is highlighted recently  \cite{Gillespie07, Kaern05, Shahrezaei08}.

Chemical dynamics have been widely accepted to study chemical kinetics of reacting systems with large molecule populations, typically in the order of $10^{23}$. In these systems, the kinetics are nearly deterministic and can be described by a set of ordinary differential equations--the \textit{reaction rate equations}, or partial differential equations if spatial movements are taken into account. Nevertheless, in intracellular molecule kinetics, stochasticity is significant because the numbers of each molecule species are very low. For instance, only one gene, either active or inactive, is involved in most activities of gene expressions. In an individual bacteria, there are less than 20 transcriptions of mRNAs from a single gene  \cite{Golding05}. Typical molecule numbers of the same protein specie in a cell are usually no more than a few thousands. Thus, fluctuations in protein activities are significant due to the low-number effect. Reaction rate equations fail to describe these fluctuations. In this review, I will introduce main equations for the stochastic modeling of such systems, and discuss numerical methods used in  stochastic simulations.

This paper will first briefly review assumptions and notations for describing a biochemical system, followed by two examples of biological processes. Next, I will introduce several theoretical formulations for modeling biochemical system kinetics with merely intrinsic noise, followed by the mathematical formulations for the situation in which kinetic parameters are random. Finally, some stochastic simulation methods  supported by the previous theories are introduced. The paper will be concluded with a summarization of the theoretical structure of stochastic modeling.

\section{Stochasticity in biological processes}

It is no doubt that biological processes are essentially random  \cite{ Fiering00, Gillespie07, Li08, McAdams97, McAdams99, Raj08, Rao02, Samoilov06, Shahrezaei08}. Both cellular behavior and the cellular environment are stochastic. Phenotypes vary across isogenic populations and in individual cell over time  \cite{Shahrezaei08}.  Gene expression is a fundamentally stochastic process, with randomness in transcription and translation leading to cell-to-cell variations in mRNA and protein levels  \cite{Elowitz02, Kaern05, Paulsson04}. Noise propagation in gene networks has important consequences for cellular functions, being beneficial in some contexts and harmful in others  \cite{Pedraza05, Raj08}. 

Essentially, kinetics of biological molecules in living cells are consequences of chemically reacting systems. In this section, we first review basic assumptions and descriptions of general chemical systems, and then give two examples to demonstrate their applications.

\subsection{Chemical systems}

Consider a system of well-stirred mixture of $N\geq 1$ molecular species $\{S_1,\cdots, S_N\}$, inside some fixed volume $\Omega$ and at constant temperature, through $M\geq 1$ reaction channels $\{R_1,\cdots, R_M\}$. We can specify the dynamical state of this system by $\vec{X}(t) = (X_1(t),\cdots, X_N(t))$, where
\begin{equation}
 \begin{array}{rcl}
X_i(t) &=& \mathrm{the\ number\ of\ } S_i \mathrm{\ molecule\ in}\\
&&{}\mathrm{the\ system\ at\ time\ } t,\quad (t=1,\cdots, N). 
\end{array}
\end{equation}
We will described the evolution of $\vec{X}(t)$ from some given initial state $\vec{X}(t_0) = \vec{x}_0$. It is obvious that $\vec{X}(t)$ is a stochastic process, because the time at which a particular reaction occurs is random. Therefore, instead of tracking a single pathway, our goal is to study the evolution of the statistical properties of system states. 

In this review paper, we always assume that each reaction, once occur, completes instantaneously. This is to be distinguished with the systems involve reactions with delay  \cite{ScWi08}. Further, we assume that the system is well stirred such that at any moment, each reactions occur with equal probability at any position. Under these assumptions, each reaction channel $R_j$ associates with a \textit{propensity function} $a_j$ and a \textit{state-change vector } $\vec{v}_j = (v_{j1},\cdots, v_{jN})$, which are defined such that
\begin{equation}
\label{eq:prop}
\begin{array}{rcl}
a_j(\vec{x}) d t &=& \mathrm{the\ probability,\ given\ } \vec{X}(t) = \vec{x}, \mathrm{\ that\ one\ reaction\ } R_j\\
&& {}\mathrm{will\ occur\ somewhere\ inside\ } \Omega \mathrm{\ in\ the\ next\ infinitesimal\ }\\
&&{} \mathrm{time\ interval\ } [t,t+dt),\quad (j=1,\cdots,M).
\end{array}
\end{equation}
and
\begin{equation}
\begin{array}{rcl}
 v_{ji} &=& \mathrm{the\ change\ in\ the\ number\ of\ } S_i \mathrm{\ molecule\  produced\ }\\
&&{}\mathrm{by\ one\ } R_j \mathrm{\ reaction},\quad (j=1,\cdots,M; i=1,\cdots N).
\end{array}
\end{equation}
The propensity function and the state-change vector together specify the reaction channel $R_j$. Therefore, the equations given below to describe the evolution of a biochemical system are derived from the propensity functions and stage-change vectors connected to the $M$ reaction channels.

The state-change vector of a reaction channel is easy to be obtained by counting the numbers of each molecule species that are consumed and produced in one reaction. For instance, if $R_1$ were the reaction $S_1 + 2 S_2 \to 2 S_3$, then $\vec{v}_1 = (-1, -2, 2,\cdots)$. 
 
\begin{figure}[htbp]
 \centering
\includegraphics[width=6cm]{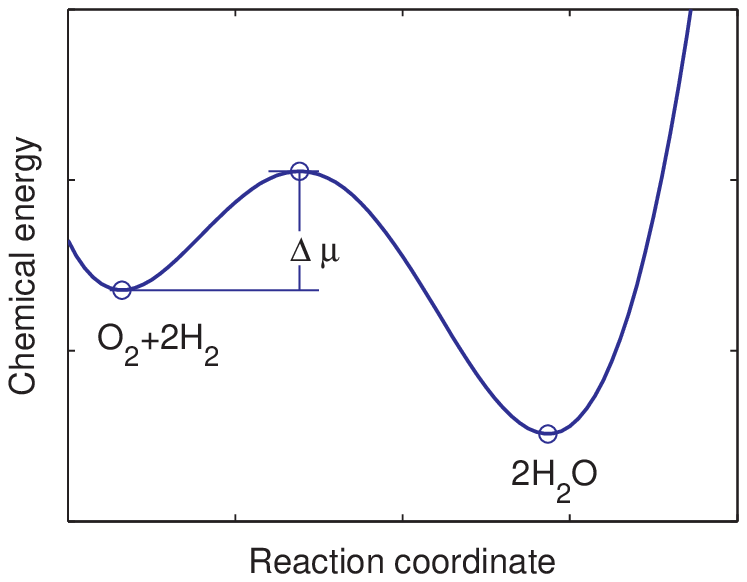}
\caption{Schematic of chemical energy for the reaction $\mathrm{O}_2 + 2 \mathrm{H}_2 \to 2 \mathrm{H}_2 \mathrm{O}$.}
\label{fig:1}
\end{figure}

Exact descriptions of propensity functions associate with the \textit{ad hoc} stochasticity of deterministic chemical kinetics  \cite{Oppenheim69}, and have solid microphysical basis. In general, the function $a_j$ have the mathematical form \cite{Gillespie:2000}
\begin{equation}
 \label{eq:aj}
a_j(\vec{x}) = c_j h_j(\vec{x}).
\end{equation}
Here $c_j$ is the \textit{specific probability rate constant} for the channel $R_j$, which is defined such that $c_j d t$ is the probability that a randomly chosen combination of $R_j$ reactant molecules will react accordingly in the next infinitesimal time interval $dt$. This probability $c_j dt$ equals the multiple of two parts, the probability of a randomly chosen combination of $R_j$ reactant molecules will collide in the next $dt$, and the probability that a colliding reactant molecules will actually react according to $R_j$. The first probability depends on the average relative speeds (which in turn depends on the temperature), the collision sections of a reactant molecules, and the system volume $\Omega$. The second probability depends on the chemical energy barrier $\Delta \mu$ of the reaction $R_j$ (Figure \ref{fig:1}), and usually associate with temperature through a Boltzmann factor $e^{-\Delta \mu/k_B T}$, where $k_B$ is the Boltzmann constant. 

The function $h_j(\vec{x})$ in \eqref{eq:aj} measures the number of distinct combinations of $R_j$ reactant molecules available in the state $\vec{x}$. It can be easily obtained from the reaction $R_j$. For example, in the above reaction $R_1$, we would have $h_j(\vec{x}) = x_1 x_2 (x_2-1)/2$, which give the number of combinations to select one $S_1$ molecule from $x_1$ of them, and two $S_2$ molecules from $x_2$ of them. More examples are given below.

In general, for a chemical reaction 
\begin{equation}
 R_j: m_{j1} S_1 + \cdots + m_{jN} S_N \to n_{j1} S_1 + \cdots + n_{jN} S_N,
\end{equation}
we would have
$$v_{ji} = n_{ji} - m_{ji},\quad a_j(\vec{x}) \propto \prod_{k=1}^N \dfrac{x_k!}{m_{jk}! (x_k - m_{jk})!}.$$
If for any $k$ and $j$, have $x_k\gg m_{jk}$, then approximately
$$a_j(\vec{x}) = c_j \prod_{k=1}^N x_k^{m_{jk}}.$$
The \textit{reaction rate constant} $c_j$ can only be obtained from experiments, and usually depends on the system volume $\Omega$ and the temperature. 

In real systems, most reaction channels are either \textit{monomolecular} or \textit{bimolecular} reaction. For a monomolecular reaction, the reaction rate constant $c_j$ is independent of the system volume $\Omega$. For a bimolecular reaction, the rate constant $c_j$ is inversely proportional $\Omega$. \textit{Trimolecular} reactions do not physically occur in dilute solutions with appreciable frequency. One can consider a trimolecular reaction as the combined result of two bimolecular reactions, and involved an additional short-lived species. For such an ``effective trimolecular'' reaction, the approximate $c_j$ is proportional to $\Omega^{-2}$  \cite{Gillespie:2000}. 

\subsection{Examples in gene regulations}

\subsubsection{Gene expression}

\begin{figure}[htbp]
 \centering
\begin{minipage}{5cm}
\includegraphics[width=5.0cm]{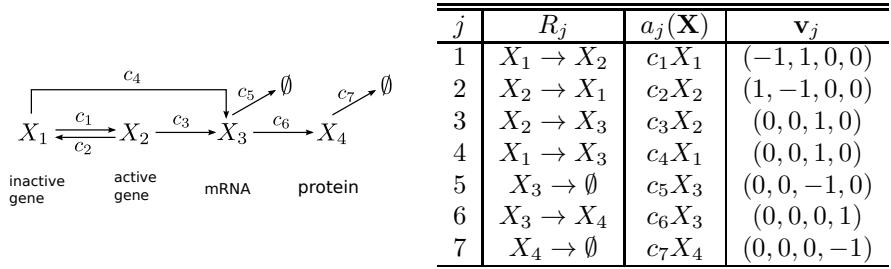}
\end{minipage}\hspace{0.5cm}
\begin{minipage}{5.5cm}
\begin{tabular}{c|c|c|c}
\hline
\hline
$j$ & $R_j$ & $a_j(\vec{X})$ & $\vec{v}_j$ \\
\hline
1 & $X_1 \rightarrow X_2$ &$c_1 X_1$ & $(-1, 1, 0, 0)$\\
2 & $X_2\rightarrow X_1$ & $c_2 X_2$ & $(1, -1, 0,0)$\\
3 & $X_2\rightarrow X_3 $ &$c_3 X_2$ & $(0, 0, 1, 0)$\\
4 & $X_1\rightarrow X_3$ &$c_4 X_1 $ & $(0,0,1,0)$\\
5 &$X_3\rightarrow \emptyset$ &$c_5 X_3 $ & $(0,0,-1,0)$\\
6 &$X_3\rightarrow X_4$ &$c_6 X_3$ & $(0,0,0,1)$\\
7 &$X_4\rightarrow \emptyset$ &$c_7 X_4$ & $(0,0,0,-1)$\\
\hline
\hline 
\end{tabular} 
\end{minipage}

\caption{ (From ref.  \cite{Lei09}) A model of gene expression. Each step represents the biochemical reactions which associate with transition between promoter states, production and decaying of mRNAs and proteins (here $c_3>c_4$).}
\label{fig:ge}
\end{figure}

Gene expression is a basic biological process. Reactions in gene expression include promoter activity and inactivity, \textit{transcription}, \textit{translation}, and decaying of mRNA and proteins. Typical steps in gene expression are illustrated in Figure \ref{fig:ge} (also refer  \cite{Lei09}). Note that transcription is a process of mRNA synthesis as specified by the gene, in which only the information is read out and the gene is not consumed. Similarly, proteins are translated from a mRNA sequence according to the \textit{genetic code}, and the mRNA is not consumed in this process. 

In many gene regulations, the transition between active and inactive promoter states are regulated by a proteins (activator or repressor). In the case of activation, the activator bind to the inactive promoter to enhance the gene expression. The reaction channels $R_1$ and $R_2$ become
$$R_1: X_5 + X_1\rightarrow X_2,\quad R_2: X_2\rightarrow X_1 + X_5,$$
where $X_5$ stands for the number of the activator. The corresponding propensity functions and state-change vectors are $a_1(\vec{X}) = c_1 X_1 X_5$,$a_2(\vec{X}) = c_2 X_2$, $\vec{v}_1 = (-1,1,0,0,-1)$, and $\vec{v}_2 = (1,-1,0,0,1)$. Similarly, in the case of repressor, we should have
$$R_1: X_1\rightarrow X_2 + X_5,\quad R_2: X_5 + X_2\rightarrow X_1,$$
and $a_1(\vec{X}) = c_1 X_1$, $a_2(\vec{X}) = c_2 X_2 X_5$, $\vec{v}_1 = (-1,1,0,0,1)$, and $\vec{v}_2 = (1,-1,0,0,-1)$.

\subsubsection{Circadian oscillator}

\begin{figure}[htbp]
 \centering
\begin{minipage}{4.5cm}
\includegraphics[width=4.50cm]{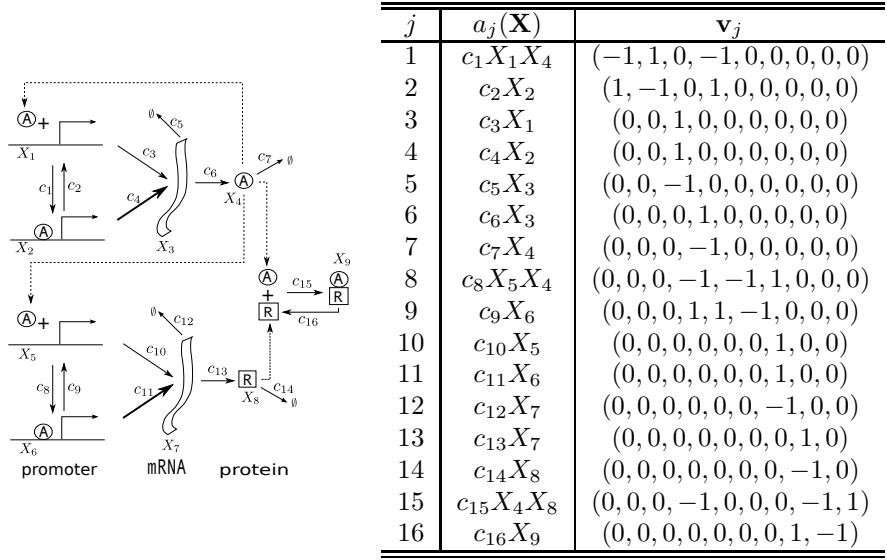}
\end{minipage}\hspace{0.3cm}
\begin{minipage}{6.0cm}
\begin{tabular}{c|c|c}
\hline
\hline
$j$ & $a_j(\vec{X})$ & $\vec{v}_j$ \\
\hline
1 & $c_1 X_1 X_4$ & $(-1, 1, 0, -1, 0, 0, 0, 0, 0)$\\
2 & $c_2 X_2$ & $(1, -1, 0,1,0,0,0,0,0)$\\
3 & $c_3 X_1$ & $(0, 0, 1, 0,0,0,0,0,0)$\\
4 & $c_4 X_2 $ & $(0,0,1,0,0,0,0,0,0)$\\
5 & $c_5 X_3 $ & $(0,0,-1,0,0,0,0,0,0)$\\
6 &$c_6 X_3$ & $(0,0,0,1,0,0,0,0,0)$\\
7 &$c_7 X_4$ & $(0,0,0,-1,0,0,0,0,0)$\\
8 & $c_8 X_5 X_4$ & $(0,0,0,-1,-1,1,0,0,0)$\\
9 & $c_9 X_6$ & $(0,0,0,1,1,-1,0,0,0)$\\
10 & $c_{10} X_5$ & $(0,0,0,0,0,0,1,0,0)$\\
11 & $c_{11} X_6$ & $(0,0,0,0,0,0,1,0,0)$\\
12 & $c_{12}X_7$ & $(0,0,0,0,0,0,-1,0,0)$\\
13 & $c_{13} X_7$ & $(0,0,0,0,0,0,0,1,0)$\\
14 & $c_{14} X_8$ & $(0,0,0,0,0,0,0,-1,0)$\\
15 & $c_{15} X_4 X_8$ & $(0,0,0,-1,0,0,0,-1,1)$\\
16 & $c_{16} X_9$ & $(0,0,0,0,0,0,0,1,-1)$ \\
\hline
\hline 
\end{tabular} 
\end{minipage}

\caption{ (From ref.  \cite{Vilar02}) A biochemical network of the circadian oscillator model.}
\label{fig:co}
\end{figure}

Figure \ref{fig:co} shows a simple model of circadian oscillator based on a common positive and negative control elements found experimentally  \cite{Vilar02}. In this model, two genes, an activator $A$ and a repressor $R$, are transcribed into mRNA and subsequently translated into protein. The activator $A$ binds to both $A$ and $R$ promoters to increase their transcription rates. The protein $R$ binds to and sequester the protein $A$, and therefore acts as a repressor. Figure \ref{fig:co} shows the propensity functions, and stage-change vectors of this model ($c_i$ in the Figure gives the rate constant of the reaction channel $R_i$). Note that in the reaction channel $R_{16}$, the complex breaks in to $R$ because of the degradation of $A$. Thus $R_{16}$ is not the reversion process of $R_{15}$.

\section{Mathematical formulations--intrinsic noise}

First, we assume that the propensity functions are time independent, i.e., the reaction rates $c_j$ in \eqref{eq:aj} are constants. In this situation, the fluctuations in the system are inherent to the system of interest (\textit{intrinsic noise}). The opposite case is \textit{extrinsic noise}, which arises from variability in factors that are consider to be external. Mathematical formulations for extrinsic noise will be discussed in next section. 

\subsection{Chemical master equation}

From the propensity function given in previous, the state vector $\vec{X}(t)$ is a jump-type Markov process on the non-negative $N$-dimensional integer lattice. In following analysis of such a system, we will focus on the conditional \textit{probability function} 
\begin{equation}
 P(\vec{x},t| \vec{x}_0, t_0) = \mathrm{Prob}\{\vec{X}(t) = \vec{x},\mathrm{\ given\ that\ } \vec{X}(t_0) = \vec{x}_0\}.
\end{equation}
Hereinafter, we use an upper case letter to denote a random variable, and the corresponding lower case letter for a possible value of that random variable. 

Through the probability function, the \textit{average} or \textit{expectation value} of any quantity $f(\vec{X}|\vec{x}_0,t_0)$ defined on the system is given by
\begin{equation}
 \langle f(\vec{X}|\vec{x}_0,t_0)\rangle = \sum_{\vec{x}} f(\vec{x}) P(\vec{x},t| \vec{x}_0, t_0).
\end{equation}

Physically, in an ensemble of identical systems starting from the same initial state $\vec{X}(t_0) = \vec{x}_0$, the function $P(\vec{x},t| \vec{x}_0, t_0)$ gives the fraction of subsystems with state $\vec{X}(t) = \vec{x}$.

To derive a time evolution for the probability function $P(\vec{x}, t|\vec{x}_0, t_0)$, we take a time increment $dt$ and consider the variation between the  probability of $\vec{X}(t) = \vec{x}$ and of $\vec{X}(t+dt) = \vec{x}$, given that $\vec{X}(t_0) = \vec{x}_0$. This variation is 
\begin{equation}
 \label{eq:cme1}
\begin{array}{rcc}
P(\vec{x},t+dt|\vec{x}_0,t_0) - P(\vec{x},t|\vec{x}_0,t_0)& =&  \mathrm{Increasing\ of\ the\ probability\ in\ } dt  \\
&&{} - \\
&&{}\mathrm{Decreasing\ of\ the\ probability\ in\ } dt
\end{array}
\end{equation}
We take $dt$ so small such that the probability of having two or more reactions in $dt$ is negligible compared to the probability of having only one reaction. Then increasing of the probability in $dt$ occurs when a system with state $\vec{X}(t) = \vec{x}-\vec{v}_j$ reacts according to $R_j$ in $(t,t+dt)$, the probability of which is $a_j(\vec{x}-\vec{v}_j)dt$. Thus, 
\begin{equation}
\label{eq:cme2}
\mathrm{Increasing\ of\ the\ probability\ in\ } dt = \sum_{j=1}^M P(\vec{x}-\vec{v}_j,t|\vec{x}_0,t_0)a_j(\vec{x}-\vec{v}_j)dt.
\end{equation}
Similarly, when a system with state $\vec{X}(t) = \vec{x}$ reacts according to any reaction channel $R_j$ in $(t,t+dt)$, the probability $P(\vec{x},t|\vec{x}_0,t_0)$ will decrease. Thus,
\begin{equation}
\label{eq:cme3}
\mathrm{Decreasing\ of\ the\ probability\ in\ } dt = \sum_{j=1}^M P(\vec{x},t|\vec{x}_0,t_0)a_j(\vec{x})dt.
\end{equation}
Substituting \eqref{eq:cme2} and \eqref{eq:cme3} into \eqref{eq:cme1}, we obtain
\begin{eqnarray*}
 P(\vec{x},t+dt|\vec{x}_0,t_0) - P(\vec{x},t|\vec{x}_0,t_0) &=&\sum_{j=1}^M P(\vec{x}-\vec{v}_j,t|\vec{x}_0,t_0)a_j(\vec{x}-\vec{v}_j)dt\\
&&{} - \sum_{j=1}^M P(\vec{x},t|\vec{x}_0,t_0)a_j(\vec{x})dt,
\end{eqnarray*}
which yields, with the limit $dt\to0$, the \textit{chemical master equation} \cite{Gillespie:2000, vanKampen}:
\begin{equation}
\label{eq:cme}
 \dfrac{\partial\ }{\partial t}P(\vec{x},t|\vec{x}_0,t_0) = \sum_{j=1}^M [P(\vec{x}-\vec{v}_j,t|\vec{x}_0,t_0)a_j(\vec{x}-\vec{v}_j) - P(\vec{x},t|\vec{x}_0,t_0)a_j(\vec{x})].
\end{equation}

The equation \eqref{eq:cme} is an exact consequence from the reaction channels characterized by propensity functions and stage-change vectors. If one can solve \eqref{eq:cme} for $P$, we should be able to find out everything about the process $\vec{X}(t)$. However, such an exact solution of \eqref{eq:cme} can rarely be obtained (refer  \cite{JaHu07, ShahrezaeiSwain} for the examples of solving the chemical master equation analytically). 

In fact, the chemical master equation \eqref{eq:cme} is a set of linear differential equations with constant coefficients. The difficult on solving the equation \eqref{eq:cme} comes from the high dimension, which equals the total number of possible states of the system under study. For example, for a system with 100 molecules species, each has two possible states ($X_i(t) = 0$ or $1$), the system has totally $2^{100}$ possible states, and therefore the equation \eqref{eq:cme} constants $2^{100}$ equations! Even solving such a huge system numerically is a big challenge.

\subsection{Fokker-Plank equation}

If in a biochemical system, all components of $\vec{X}(t)$ are very large compared to $1$, we can regard the components of $\vec{X}(t)$ as real numbers. We assume further that the functions $f_j(\vec{x}) \equiv a_j(\vec{x}) P(\vec{x}, t|\vec{x}_0, t_0)$ are analytic in the variable $\vec{x}$. With these two assumptions, we can use Taylor's expansion to write
\begin{equation}
\label{eq:fp1}
f_j(\vec{x}-\vec{v}_j)=f_j(\vec{x}) + \sum_{|\vec{m}|\geq 1} \prod_{i=1}^N \dfrac{(-1)^{m_i}}{m_i!}\left(\dfrac{\partial\ }{\partial x_i}\right)^{m_i}(v_{ji}^{m_i}f_j(\vec{x})).
\end{equation}
Here $\vec{m} = (m_1,\cdots, m_N)\in \mathbb{Z}^N, |\vec{m}| = m_1 + \cdots + m_N$. 
Substituting \eqref{eq:fp1} into \eqref{eq:cme}, we immediately obtain the 
\textit{chemical Kramer-Moyal equation} (CKME)  \cite{Gillespie:2000}
\begin{equation}
 \label{eq:fp3}
\dfrac{\partial\ }{\partial t}P(\vec{x},t) = \sum_{|\vec{m}|\geq 1} \prod_{i=1}^N \dfrac{(-1)^{m_i}}{m_i!}\left(\dfrac{\partial\ }{\partial x_i}\right)^{m_i}\left(A^{m_1,\cdots,m_N}(\vec{x}) P(\vec{x},t)\right),
\end{equation}
where 
$$A^{m_1,\cdots,m_N}(\vec{x}) = \sum_{j=1}^M v_{j1}^{m_1}\cdots v_{jN}^{m_N}a_j(\vec{x}).$$
Hereinafter, we omit the initial condition $(\vec{x}_0, t_0)$. If all functions $f_j(\vec{x}-\vec{v}_j)$ are smooth, the equation \eqref{eq:fp3} would equivalent to \eqref{eq:cme}. Therefore, the chemical Kramer-Moyal equation is a ``semi-rigorous'' consequence of the chemical master equation \eqref{eq:cme}.

If we truncate the right hand side at $|\vec{m}| = 2$, we obtain the  \textit{chemical Fokker-Plank equation} (CFPE)
\begin{equation}
 \label{eq:fp2}
\dfrac{\partial\ }{\partial t}P(\vec{x},t) = - \sum_{i=1}^N \dfrac{\partial\ }{\partial x_i}A_i(\vec{x}) P(\vec{x},t) + \dfrac{1}{2}\sum_{1\leq i, k\leq N}\dfrac{\partial^2 }{\partial x_i \partial x_k}B_{ik}(\vec{x})P(\vec{x}, t).
\end{equation}
where
\begin{equation}
\label{eq:fp4}
 A_i(\vec{x}) = \sum_{j=1}^Mv_{ji} a_j(\vec{x}),\quad B_{ik}(\vec{x}) = \sum_{j=1}^M v_{ji}v_{jk} a_j(\vec{x}).
\end{equation}
We will re-obtain this chemical Fokker-Plank equation below from the chemical Langevin equation.

\subsection{Reaction rate equation}

If we multiply the chemical master equation \eqref{eq:cme} through by $x_i$, and sum over all $\vec{X}$, we obtain the following \textit{chemical ensemble average equation} (CEAE)
\begin{equation}
 \label{eq:aveX}
\dfrac{d \langle X_i\rangle}{d t} = \sum_{j=1}^M v_{ji}\langle a_j (\vec{X})\rangle \quad (i = 1,2,\cdots, N).
\end{equation}

Equation \eqref{eq:aveX} is an exact consequence of \eqref{eq:cme}. Nevertheless, \eqref{eq:aveX} is not a close equation system unless all reaction channels are monomolecular. When all reaction channels are monomolecular, the propensity functions $a_j(\vec{X})$ are linear, and therefore
\begin{equation}
\label{eq:c0}
 \langle a_j(\vec{X})\rangle = a_j(\langle \vec{X}\rangle)\quad (j=1,\cdots, N).
\end{equation}
Thus, the populations evolve deterministically according to a set of ordinary differential equations
\begin{equation}
 \label{eq:cre}
\dfrac{d x_i}{d t} = \sum_{j=1}^M v_{ji}a_j(\vec{x})\quad (i=1,\cdots, N),
\end{equation}
where the components of $\vec{x}(t)$ are now considered as real variables.

In usual, the mean populations do not evolve according to \eqref{eq:cre} when there are higher order reactions. However, \eqref{eq:cre} is sometimes heuristic if we assuming that the fluctuations  are not important and \eqref{eq:c0} holds approximately. Consequently, the ordinary differential equations \eqref{eq:cre} also hold under this assumption. The equation \eqref{eq:cre} is referred as the macroscopic \textit{reaction rate equation}(RRE), or \textit{chemical rate equation} in some literatures. 

The reaction rate equation \eqref{eq:cre} is often written in terms of the species concentrations
$$Z_i(t) \equiv X_i(t)/\Omega\quad (i=1,\cdots, N).$$
The ``concentrations'' form of the reaction rate equation has form
\begin{equation}
 \label{eq:cre1}
\dfrac{d  z_i}{d t} = \sum_{j=1}^M v_{ji}\tilde{a}_j(\vec{z})\quad (i=1,\cdots, N),
\end{equation}
where
$$\tilde{a}_j(\vec{z}) = a_j(\Omega \vec{z})/\Omega.$$
While we examine the $\Omega$ dependence in the propensity functions for monomolecular, bimolecular, and trimolecular reactions, we would find that the function $\tilde{a}_j$ is functionally identical to $a_j$ except that the rate constants $c_j$ has been replaced by the \textit{reaction rate constants} $k_j$.

The reaction rate equations are most commonly used in modeling biochemical reacting systems. However, this equation fails to describe the stochastic effects which can be very important in biological processes. We will introduce the \textit{chemical Langevin equation} below that has a facility to describe the stochasticity, and easy to be studied, at least numerically.

\subsection{Chemical Langevin equation}

The chemical Langevin equation was derived to yield an approximate time-evolution equation of the Langevin type. The derivation of the equation, given by Gillespie, is based on the chemical master equation and two explicit dynamical conditions as detailed below. Most of the following refer to Gillespie's original paper  \cite{Gillespie:2000}.

Suppose that the state of a system at current time $t$ is $\vec{X}(t) = \vec{x}$. Let $K_j(\vec{x},\tau)\ (\tau > 0)$ be the number of $R_j$ reactions that occur in the subsequent time interval $[t, t+\tau]$. Since each  of these reactions will change the $S_i$ population by $v_{ji}$, the number of $S_i$ molecule in the system at time $t+\tau$ will be
\begin{equation}
\label{eq:x1}
 X_i(t+\tau) = x_i + \sum_{j=1}^M K_j(\vec{x},\tau) v_{ji},\quad (i=1,\cdots, N).
\end{equation}
We note that $K_j(\vec{x},\tau)$ is a \textit{random variable}, and therefore $X_i(t+\tau)$ is random. 

We will obtain an approximation of $K_j(\vec{x},\tau)$ below by imposing the following conditions:
\begin{description}
 \item[Condition (i)] Require $\tau$ to be \textit{small} enough such that the change in the state during $[t,t+\tau]$ will be so slight that non of the propensity functions changes its value ``appreciably''.
\item[Condition (ii)] Require $\tau$ to be \textit{large} enough that the expected number of occurrences of each reaction channel $R_j$ in $[t, t+\tau]$ be much larger than $1$.
 \end{description}

From condition (i), the propensity functions satisfy
\begin{equation}
 a_j(\vec{X}(t')) \approx a_j(\vec{x}), \quad \forall t'\in [t,t+\tau], \forall j\in [1,M].
\end{equation}
Thus, the probability of the reaction $R_j$ to occur in any infinitesimal interval $d\tau$ within $[t, t+\tau]$ is $a_j(\vec{x}) d \tau$. Thus, $K_j(\vec{x},\tau)$,  the occurrence of ``events'' of reaction $R_j$ in the time interval $[t, t+\tau]$, will be a statistically independent \textit{Poisson} random variable, and is denoted by $\mathcal{P}_j(a_j(\vec{x}), \tau)$. So \eqref{eq:x1} can be approximated by
\begin{equation}
 \label{eq:x2}
X_i(t+\tau) = x_i + \sum_{j=1}^M v_{ji} \mathcal{P}_j(a_j(\vec{x}), \tau),\quad (i=1,\cdots, N)
\end{equation}
according to condition (i). 

The mean and variance of $\mathcal{P}(a, \tau)$ are
\begin{equation}
 \langle \mathcal{P}(a, \tau)\rangle = \mathrm{var}\{\mathcal{P}(a, \tau)\} = a \tau.
\end{equation}
Thus, condition (ii) means
\begin{equation}
\label{eq:x3}
 a_j(\vec{x})\tau \gg 1,\quad \forall j \in [1,M].
\end{equation}
The inequality \eqref{eq:x3} allows us to approximate each Poisson random variable by a \textit{normal} random variable with the same mean and variance. This leads to further approximation
\begin{equation}
 \label{eq:x4}
X_i(t+\tau) = x_i + \sum_{j=1}^M v_{ji} \mathcal{N}_j(a_j(\vec{x})\tau, a_j(\vec{x})\tau),\quad (i=1,\cdots, N)
\end{equation}
where $\mathcal{N}_j(m, \sigma^2)$ denotes the normal random variable with mean $m$ and variance $\sigma^2$. Here the $M$ random variables are independent to each other. Notice that in the above approximation, we have converted the molecular population $X_i$ from discretely changing integers to continuously changing real  variables.

We note the linear combination theorem for normal random variables,
\begin{equation}
 \mathcal{N}(m,\sigma^2) = m + \sigma \mathcal{N}(0,1),
\end{equation}
the equation \eqref{eq:x4} can be rewritten as
\begin{equation}
\label{eq:x5}
X_i(t+\tau) = x_i + \sum_{j=1}^M v_{ji}a_j(\vec{x}) \tau + \sum_{j=1}^M v_{ji}\sqrt{a_j(\vec{x})\tau} \mathcal{N}_j(0,1)\quad (i=1,\cdots, N).
\end{equation}

Now, we are ready to obtain Langevin type equations by making some purely notational changes. First, we denote the time interval $\tau$ by $dt$, and write
$$d X_i = X_i(t+dt) - X_i(t).$$
Next, introduce $M$ temporally uncorrelated, independent random process $W_j(t)$, satisfying
\begin{equation}
dW_j(t) = W_j(t+dt) - W_j(t) = \mathcal{N}_j(0,1)\sqrt{dt}\quad (j=1,\cdots, N). 
\end{equation}
It is easy to verify that the processes $W_j(t)$ have \textit{stationary independent increments with mean 0}, i.e.,
\begin{equation}
\langle d W_j(t)\rangle = 0,\ \langle d W_i(t) dW_j(t') \rangle = \delta_{ij}\delta(t-t')dt,\quad \forall 1\leq i,j\leq M, \forall t, t'.
\end{equation}
Thus, each $W_j$ is a \textit{Wiener process} (or referred to as \textit{Brownian motion}) \cite{vanKampen}. Finally, recalling that $\vec{x}$ stands for $\vec{X}(t)$, the equation \eqref{eq:x4} becomes a \textit{Langevin equation} (or \textit{stochastic differential equation})
\begin{equation}
 \label{eq:cle}
d X_i = \sum_{j=1}^M v_{ji} a_j(\vec{X}) d t + \sum_{j=1}^M v_{ji} \sqrt{a_j(\vec{X})} d W_j\quad (i=1,\cdots, N).
\end{equation}
The stochastic differential equation \eqref{eq:cle} is the desired \textit{chemical Langevin equation} (CLE). The solution of \eqref{eq:cle} with initial condition $\vec{X}(0) = \vec{X}_0$ is a stochastic process $\vec{X}(t)$ satisfying
\begin{equation}
 \vec{X}(t) = \vec{X}_0 + \sum_{j=1}^M \int_0^tv_{ji} a_j(\vec{X}(s)) d s +  \sum_{j=1}^M \int_0^tv_{ji} \sqrt{a_j(\vec{X}(s))} d W_j(s).
\end{equation}
Here \textit{It\^{o} integral} is used as the stochastic fluctuations are \textit{intrinsic} \cite{vanKampen}.

In the above, it is obvious that the conditions (i) and (ii) are contradict to each other. It may be very well happen that both conditions cannot be satisfied simultaneously. In this case, the derivation of the chemical Langevin equation may fail. But there are many practical circumstances in which the two conditions can be simultaneously satisfied. As the inequality \eqref{eq:x3} implies that $a_j(\vec{x})$ is large when $\tau$ is small enough as required by condition (i). This is possible when the system has large molecular populations for each molecule species since $a_j(\vec{x})$ is typically proportional to one or more components of $\vec{x}$. 

Even when the conditions (i) and (ii) cannot be satisfied simultaneously, the chemical Langevin equation \eqref{eq:cle} is still useful. As we will discuss below.

\subsection{Discussions about the chemical Langevin equation}

In many intracellular biochemical systems such as gene expressions and genetic networks, the molecule populations are small, and the reactions are usually slow. In these systems, the derivation of the chemical Langevin equation may fail. Nevertheless, the equation \eqref{eq:cle} is still useful for such systems as it can provide reasonable descriptions for the statistical properties of the kinetic processes. The reasons are given below.

\subsubsection{Time-evolution of the mean}

If we take average to both sides of \eqref{eq:cle}, noticing 
$$\langle \sqrt{a_j(\vec{X})}dW_j\rangle = 0 \quad (j=1,\cdots, M)$$
according to the It\^{o} interpretation, we have
\begin{equation}
 \dfrac{d \langle X_i\rangle}{dt} = \sum_{j=1}^M v_{ji} \langle a_j(\vec{X}) \rangle \quad (i=1,\cdots, N).
\end{equation}
This gives the same form of chemical ensemble average equation \eqref{eq:aveX} as we have obtained from the chemical master equation. 

\subsubsection{Time-evolution of the correlations}

The correlations between the molecule numbers are defined as
\begin{equation}
 \sigma_{ij}(t) = \langle X_i(t)X_j(t)\rangle - \langle X_i(t)\rangle \langle X_j(t)\rangle,\quad (1\leq i,j\leq N).
\end{equation}

Multiply the chemical master equation \eqref{eq:cme} through by $x_ix_j$ and sum over all $\vec{x}$, we obtain 
\begin{equation}
 \label{eq:fd1}
\dfrac{d \langle X_i X_j\rangle}{d t} = \sum_{k=1}^M \left[\langle X_i v_{kj} a_k(\vec{X}) \rangle + \langle X_j v_{ki}a_k(\vec{X})\rangle + \langle v_{ki}v_{kj}a_k(\vec{X})\rangle\right].
\end{equation}
Furthermore, \eqref{eq:aveX} gives
\begin{equation}
 \dfrac{d \langle X_i\rangle \langle X_j\rangle}{d t} = \sum_{k=1}^M \left[\langle v_{ki}a_k(\vec{X})\rangle \langle X_j\rangle + \langle v_{kj}a_k(\vec{X})\rangle\langle X_i\rangle\right].
\end{equation}
Thus, recalling \eqref{eq:fp4}, we have the \textit{chemical ensemble correlations equation} (CECE)
\begin{equation}
\label{eq:fdt1}
\dfrac{d \sigma_{ij}}{d t} = \langle A_i(\vec{X}) (X_j - \langle X_j\rangle)\rangle + \langle (X_i - \langle X_i\rangle) A_j(\vec{X})\rangle + \langle B_{ij}(\vec{X})\rangle.
\end{equation}
The equation \eqref{eq:fdt1} is \textit{exact} from the chemical master equation.

Now, starting from the chemical Langevin equation \eqref{eq:cle} and applying the It\^{o} formula  \cite{Oksendal}, we have
\begin{eqnarray*}
 d(X_i X_j) &=& X_i (d X_j) + X_j (d X_i) + (d X_i) (d X_j)\\
&=&\sum_{k=1}^M [X_j v_{ki}a_k(\vec{X}) + X_i v_{kj}a_k(\vec{X})X_i +  v_{ki}v_{kj}a_k(\vec{X})] d t\\
&&{} + \sum_{k=1}^M[X_i v_{kj}\sqrt{a_k(\vec{X})} + X_j v_{kj} \sqrt{a_k(\vec{X})}]d W_k + o(dt)
\end{eqnarray*}
Taking the average to both sides, we re-obtain \eqref{eq:fd1}, which yields the same form of chemical ensemble correlations equation \eqref{eq:fdt1}. 

At states of near equilibrium, if the random fluctuations are not important, we have approximately
$$A_i(\vec{X})\approx A_i(\langle\vec{X}\rangle) + \sum_{l=1}^N \dfrac{\partial A_i(\langle \vec{X}\rangle)}{\partial X_l}(X_l - \langle X_l\rangle),\quad \langle B_{ij}(\vec{X})\rangle\approx B_{ij}(\langle \vec{X}\rangle).$$
Substituting above approximations into \eqref{eq:fdt1}, and defining the matrices
\begin{equation}
 \sigma = (\sigma_{ij}),\quad A = \left(\dfrac{\partial A_i(\langle \vec{X}\rangle)}{\partial X_l}\right),\quad B = (B_{ij}(\langle \vec{X}\rangle)),
\end{equation}
we have
\begin{equation}
 \label{eq:fdt2}
\dfrac{d \sigma}{d t} = (A \sigma + \sigma A^T)  + B.
\end{equation}

Equation \eqref{eq:fdt2} gives the linear approximation of the time-evolution of the correlation functions near equilibrium. This an example of the well known \textit{fluctuation-dissipation theorem} \cite{vanKampen}. 

\subsubsection{Fokker-Plank equation}

Consider the chemical Langevin equation \eqref{eq:cle}, the solution $\vec{X}(t)$ satisfies
\begin{equation}
X_i(t+\Delta t) = X_i(t) + \sum_{k=1}^M v_{ki}a_j(\vec{X}) \Delta t + \sum_{k=1}^M v_{ki} \sqrt{a_k(\vec{X})} \Delta W_k(t),\quad (i=1,\cdots, N)
\end{equation}
where
$$\Delta W_k(t) = W_k(t+\Delta t) - W_k(t).$$
Thus, assuming $\vec{X}(t) = \vec{x}$, the displacement $\Delta \vec{x} = \vec{X}(t+\Delta t) - \vec{X}(t)$ satisfies
$$\langle \Delta x_i\rangle = \sum_{k=1}^M v_{ki} a_k(\vec{x}) \Delta t $$
and
$$\langle \Delta x_i \Delta x_j\rangle = \sum_{k=1}^Mv_{ki}v_{kj}a_k(\vec{x}) \Delta t.$$

Let $P(\vec{x}, t)$ to be the conditional probability density function as previous, then $P(\vec{x}, t)$ satisfies
\begin{eqnarray*}
P(\vec{x}, t+dt) - P(\vec{x},t) &=& \int_{\Delta \vec{x}\in \mathbb{R}^N} P(\vec{x}-\Delta \vec{x}, t) W(\Delta \vec{x}, dt; \vec{x}-\Delta \vec{x}, t) d \Delta \vec{x}\\
&&{} - \int_{\Delta \vec{x}\in \mathbb{R}^N} P(\vec{x}, t) W(\Delta \vec{x}, dt; \vec{x}, t) d \Delta \vec{x},
\end{eqnarray*}
where $W(\Delta \vec{x}, dt; \vec{x}, t)$ is the transition probability from $\vec{X}(t) = \vec{x}$ to $\vec{X}(t+dt) = \vec{x}+\Delta \vec{x}$. 
Applying Taylor expansion to the function 
$$h(\vec{x}-\Delta \vec{x})\equiv P(\vec{x} -\Delta \vec{x}, t) W(\Delta \vec{x}, dt; \vec{x} - \vec{\Delta x}, t),$$ 
noticing
$$\int_{\Delta \vec{x}\in \mathbb{R}^N} \Delta x_i W(\Delta \vec{x}, dt; \vec{x}, t) = \langle \Delta x_i\rangle,$$
and
$$\int_{\Delta \vec{x}\in\mathbb{R}^N} \Delta x_i \Delta_j W(\Delta \vec{x}, dt; \vec{x},t )d \Delta x = \langle \Delta x_i \Delta x_j\rangle,$$
we have
\begin{eqnarray*}
P(\vec{x}, t+dt) - P(\vec{x},t) &=& -\sum_{i=1}^N \dfrac{\partial\ }{\partial x_i}\left[P(\vec{x},t)\langle \Delta x_i\rangle\right]\\
&& + \dfrac{1}{2}\sum_{1\leq i,j\leq N} \dfrac{\partial^2\ }{\partial x_i \partial x_j} \left[P(\vec{x},t) \langle \Delta x_i \Delta x_j\rangle \right].
\end{eqnarray*}
Thus, letting $dt\to 0$ (here $\Delta t = d t$), we re-obtain the chemical Fokker-Plank equation \eqref{eq:fp2} as previous.

Thus, the chemical Langevin equation and the chemical master equation yield the same form of chemical Fokker-Plank equation.

\subsubsection{Remarks}
In comparing the chemical Langevin equation and chemical master equation, we should note the following:
\begin{enumerate}
\item In this section, we have shown that the chemical Langevin equation and chemical master equation yield the same forms of chemical ensemble average equation \eqref{eq:aveX},  chemical ensemble correlation equation \eqref{eq:fdt1}, and chemical Fokker-Plank equation \eqref{eq:fp2}.  
\item Despite the same forms of equations \eqref{eq:aveX} and \eqref{eq:fdt1}, since the averages are taken with respect to probability functions $P(\vec{x}, t)$, which are obtained from either the solutions of the chemical master equation  or the chemical Langevin equation, the two equations (CLE and CME) do not always yield the same dynamics of ensemble average and correlation.  These are possible when all reaction channels are monomolecular, in which cases the two equations \eqref{eq:aveX} and \eqref{eq:fdt1} are, as we have discussed before, close.
\item The chemical Langevin equation and chemical master equation yield the same form of chemical Fokker-Plan equation, which is a close equation. Thus, up to the second order approximation, the two equations give the same time-evolutions of the probability function $P(\vec{x},t)$.
\end{enumerate}

\section{Mathematical formulations--fluctuation in kinetic parameters}

In the previous discussion, we assume that the reaction rates $c_j$ in \eqref{eq:aj} are constants. This is only a rough approximation of the real world, in which cell environments are stochastic, and therefore the reaction rates are random. 
Here, we will discuss the mathematical formulations for situations in which there are fluctuations in kinetic parameters.

\subsection{Reaction rate as a random process}

When the reaction rate $c_j$ is random, the previous propensity function $a_j(\vec{x})$ in \eqref{eq:aj} should be rewritten as
\begin{equation}
\label{eq:ajt}
a_j(\vec{x},t) = c_j(t) h_j(\vec{x}).
\end{equation}
Here $c_j(t)$, as previous, is the specific probability reaction rate for channel $R_j$ at time $t$. This rate is usually random, depending on the fluctuating environment whose explicit time dependence is not known. A reasonable approximation is given below.

In previous discussions, we have obtained $c_j \propto e^{-\Delta \mu_j/k_B T}$. If there are noise perturbations to the energy barrier, we can replace $\Delta \mu_j$ by $\Delta \bar{\mu}_j - k_B T\eta_j(t)$, where $\eta_j(t)$ is a stochastic process. Consequently,  we can replace the reaction rate $c_j(t)$ by
\begin{equation}
\label{eq:cjt}
 c_j(t) = \bar{c}_j e^{\eta_j(t)}/\langle e^{\eta_j(t)}\rangle,
\end{equation}
where $\bar{c}_j$ is a constant, measures the mean of the reaction rate. 

In many cases, the noise perturbation $\eta(t)$ (here we omit the subscript $j$) can be described by an \textit{Ornsterin-Uhlenbeck process}, which is given by a solution of following stochastic differential equation
\begin{equation}
\label{eq:ou}
 d \eta = -(\eta/\tau) d t  + (\sigma/\tau) d W.
\end{equation}
Here $W$ is a Wiener process, $\tau$ and $\sigma$ are positive constants, measuring the autocorrelation time and variance, respectively. It is easy to verify that $\eta(t)$ is normally distributed and has an exponentially decaying stationary autocorrelation function \cite{ Gillespie92, OU:30, Kampen:89, vanKampen}
\begin{equation}
 \langle \eta(t)\eta(t')\rangle = \dfrac{\sigma^2}{2 \tau} e^{-|t-t'|/\tau}.
\end{equation}
The Ornstein-Uhlenbeck process is an example of \textit{color noise}.  When $\tau\to 0$, $\eta(t)$ approaches to \textit{Gaussian white noise}.

With $\eta(t)$ an Ornstein-Uhlenbeck process, the stationary distribution of the reaction rate $c_j(t)$ is then \textit{log-normal}. Log-normal distribution have been seen in many applications \cite{Limpert01}. For instance, log-normal rather than normal distribution have been measured for gene expression rates \cite{Rosenfeld05, Sommer}. 

When $\eta(t)$ is an Orstein-Uhlenbeck process, we have  \cite{Crow}
\begin{equation}
\langle e^{\eta(t)}\rangle = e^{\sigma^2/(4\tau)}.
\end{equation} 
Therefore, the reaction rate \eqref{eq:cjt} can be rewritten as
\begin{equation}
 c_j(t) = \bar{c}_j e^{\eta_j(t)-\sigma_j^2/(4\tau_j)}.
\end{equation}
Thus, the propensity function \eqref{eq:ajt} now becomes
\begin{equation}
\label{eq:ajt1}
 a_j(\vec{x},t) = \bar{c}_j e^{\eta_j(t)-\sigma_j^2/(4\tau_j)} h_j(\vec{x}),
\end{equation}
with $\eta_j(t)$ satisfies an equation of form \eqref{eq:ou}.

Next, we will introduce generalizations of the reaction rate equation and the chemical Langevin equation to describe reaction systems with fluctuations in reaction rates. We note that in this situation, the state variable $\vec{X}$ alone is not enough to describe the state of a system, and the state of environment has to be taken into account as well. Thus, to generalize the chemical master equation or the Fokker-Plank equation, we should replace the previous probability function with $P(\vec{x}, \vec{c}, t|\vec{x}_0, \vec{c}_0, t)$, where $\vec{c} = \{c_1,\cdots, c_M\}$. This consideration is omitted here.

\subsection{Reaction rate equation}

Substituting the propensity functions of form \eqref{eq:ajt1} and \eqref{eq:ou} into the reaction rate equation \eqref{eq:cre}, we obtain the following stochastic differential equations
\begin{eqnarray}
 \label{eq:gcre1}
d X_i &=& \left(\sum_{j=1}^M v_{ji} \bar{c}_j e^{\eta_j - \sigma_j^2/(4\tau_j)}h_j(\vec{X})\right) d t,\quad (i=1,\dots, N)\\
\label{eq:gcre2}
d \eta_j &=& -(\eta_j/\tau_j) d t + (\sigma_j /\tau_j) d W_j,\quad (j=1,\cdots, M)
\end{eqnarray}
Here $\bar{c}_j, \tau_j, \sigma_j$ are positive constants. The equations \eqref{eq:gcre1}-\eqref{eq:gcre2} generalize the reaction rate equation to describe the reaction system with fluctuations in the kinetic parameters.

\subsection{Chemical Langevin equation}

Similar to the above strategy, substituting the propensity functions of form \eqref{eq:ajt1} and \eqref{eq:ou} into the chemical Langevin equation \eqref{eq:cle}, we obtain the following stochastic differential equations
\begin{eqnarray}
 \label{eq:gcle1}
d X_i &=& \left(\sum_{j=1}^M v_{ji} \bar{c}_j e^{\eta_j - \sigma_j^2/(4\tau_j)}h_j(\vec{X})\right) d t\\
&&{} + \sum_{j=1}^M v_{ji} \left(\bar{c}_j e^{-\eta_j - \sigma_j^2/(4\tau_j)}h_j(\vec{X})\right)^{1/2}dW_j,\quad (i=1,\dots, N)\nonumber\\
\label{eq:gcle2}
d \eta_j &=& -(\eta_j/\tau_j) d t + (\sigma_j/\tau_j) d W_{j}',\quad (j=1,\cdots, M)
\end{eqnarray}
Here $W_j'$ are independent Wiener processes, and $\bar{c}_j, \tau_j, \sigma_j$, as previous, are positive constants. This set of generalized chemical Langevin equations describes the stochasticity of chemical systems with both intrinsic noise and fluctuations in kinetic parameters.

\section{Stochastic simulations}

We have introduced several equations for modeling systems of biochemical reactions. 
Here, we will introduce stochastic simulation methods that intend to mimic a random process $\vec{X}(t)$ of a system evolution. Once we have enough sampling pathways of the random process, we are able to calculate the probability density function $P(\vec{x}, t)$ and other statistical behaviors including the mean trajectory and correlations.

\subsection{Stochastic simulation algorithm}

Assume the system in state $\vec{x}$ at time $t$. We define the \textit{next-reaction density function} $p(\tau, j|\vec{x} , t)$ as
\begin{equation}
 \label{eq:nrdf}
\begin{array}{rcl}
p(\tau, j|\vec{x}, t) d \tau & = &\mathrm{probability\ that,\ given}\ \vec{X}(t) = \vec{x},\ \mathrm{the\ next}\\ 
&&{}\mathrm{reaction\ in}\ \Omega\ \mathrm{will\ occur\ in\ the\ infinitesimal\ time}\\
&&{}\mathrm{interval}\ [t+\tau, t+\tau + d \tau),\ \mathrm{and\ will\ be\ an}\\
&&{} R_j\ \mathrm{reaction}.
\end{array}
\end{equation}
An elementary probability argument based on the propensity function \eqref{eq:prop} yields  \cite{Gillespie76, Gillespie77}
\begin{equation}
\label{eq:ga}
p(\tau, j|\vec{x}, t)  = a_j(\vec{x}) \exp(- a_0(\vec{x}) \tau),\quad (0\leq \tau < \infty, j = 1,\cdots, M) 
\end{equation}
where
$$a_0(\vec{x}) = \sum_{k=1}^Ma_k(\vec{x}).$$
This formula provides the basis for the \textit{stochastic simulation algorithm}(SSA) (also known as the \textit{Gillespie algorithm}). 

Following direct method is perhaps the simplest to generate a pair of numbers $(\tau, j)$ in accordance with the probability \eqref{eq:ga} \cite{Gillespie77}: First generate two random numbers $r_1$ and $r_2$ of uniform distribution in the unit interval, and then take
\begin{eqnarray}
 \tau &=&(1/a_0(\vec{x})) \ln (1/r_1)\\
j&=&\mathrm{the\ smallest\ integer\ satisfying\ }\sum_{j'=1}^j a_{j'}(\vec{x}) > r_2 a_0(\vec{x}).
\end{eqnarray}

Below are main steps in the stochastic simulation algorithm (for details, refer  \cite{Gillespie77})
\begin{enumerate}
 \item \textbf{Initialization}: Let $\vec{X} = \vec{x}_0$, and $t = t_0$.
\item \textbf{Monte Carlo step}: Generate a pair of random numbers $(\tau, j)$ according to the probability density function \eqref{eq:ga} (replace $\vec{x}$ by $\vec{X}$).
\item \textbf{Update}: Increase the time by $\tau$, and replace the molecule count by $\vec{X} + \vec{v}_j$.
\item \textbf{Iterate}: Go back to Step 2 unless the simulation time has been exceeded.
\end{enumerate}
The stochastic simulation algorithm consists with the chemical master equation in the sense that \eqref{eq:cme} and \eqref{eq:ga} are exact consequence of the propensity function  \eqref{eq:prop}. This algorithm numerically simulation the time evolution of a given chemical system, and gives a sample trajectory of the real system.

\subsection{Tau-leaping algorithm}

In the derivation of the chemical Langevin equation, we can always select $\tau$ small enough such that the the following \textit{leap condition} is satisfied: 
\begin{center}
\begin{minipage}{10cm}
\textit{Leap condition:} During $[t, t+\tau)$, no propensity function is likely to change its value by a significant amount.
\end{minipage} 
\end{center}
Consequently, the previous arguments indicate that we can approximately leap the system with a time $\tau$ by taking 
\begin{equation}
\label{eq:leap}
 \vec{X}(t+\tau) \approx \vec{x} + \sum_{j=1}^M \mathcal{P}_j(a_j(\vec{x}), \tau)\vec{v}_j,
\end{equation}
where $\vec{x} = \vec{X}(t)$. 

The equation \eqref{eq:leap} is the basic of the \textit{tau-leaping algorithm} \cite{Gillespie01}: Starting from the current state $\vec{x}$, we first choose a value $\tau$ that satisfies the leap condition. Next, we generate for each $j$ a random number $k_j$ according to Poisson distribution with mean $a_j(\vec{x})\tau$. Finally, we update the state from $\vec{x}$ to $\vec{x}+\sum_j k_j \vec{v}_j$, and increase the time by $\tau$. 

There are two practical issues need to be resolved in order to effectively apply the tau-leaping algorithm: First, how can we estimate the largest value of $\tau$ that satisfies the leap condition? Second, how can we ensure that the generated $k_j$ values do not result in negative populations?

The original estimation of the largest value of $\tau$ was given by Gillespie and is sketched below  \cite{Gillespie01}.  If $\tau$ satisfies the leap condition, from \eqref{eq:leap}, the average state changes over time $\tau$ is 
$$\bar{\vec{\lambda}} = \sum_{j=1}^M \langle \mathcal{P}_j(a_j(\vec{x}),\tau)\rangle\vec{v}_j = \tau \vec{\xi}(\vec{x}),$$
where
\begin{equation}
\vec{\xi}(\vec{x}) = \sum_{j=1}^M a_j(\vec{x}) \vec{v}_j
\end{equation}
is the state change per unit time. Therefore, the average difference in the propensity function $a_j(\vec{x})$ is given by $|a_j(\vec{x} +\bar{\vec{\lambda}}) - a_j(\vec{x})|$, which, according to the leap condition, should satisfies
\begin{equation}
\label{eq:lp2}
|a_j(\vec{x} +\bar{\vec{\lambda}}) - a_j(\vec{x})| \leq \varepsilon a_0(\vec{x}),\quad (j = 1,\cdots, M),
\end{equation}
where $0<\varepsilon\ll 1$, and
\begin{equation}
a_0(\vec{x}) = \sum_{j=1}^Ma_j(\vec{x}).
\end{equation}
Since
$$|a_j(\vec{x} +\bar{\vec{\lambda}}) - a_j(\vec{x})| \approx \bar{\vec{\lambda}} \cdot \nabla a_j(\vec{x}) = \sum_{i=1}^N \tau \vec{\xi}_i (\vec{x}) \dfrac{\partial a_j(\vec{x})}{\partial x_i}.$$
Let
\begin{equation}
b_{ji}(\vec{x}) = \dfrac{\partial a_j(\vec{x})}{\partial x_i},
\end{equation}
\eqref{eq:lp2} can be approximated by
$$\tau |\sum_{j=1}^N \vec{\xi}_i(\vec{x}) b_{ji}(\vec{x})|\leq \varepsilon a_0(\vec{x}),\quad (j=1,\cdots, M),$$
which yields
$$\tau \leq \varepsilon a_0(\vec{x})/|\sum_{j=1}^N \vec{\xi}_i(\vec{x}) b_{ji}(\vec{x})|.$$
Thus, the largest value of $\tau$ is given by
\begin{equation}
 \tau = \min_{j\in [1, M]} \left\{\varepsilon a_0(\vec{x})/|\sum_{i=1}^N\vec{\xi}_i(\vec{x})b_{ji}(\vec{x})|\right\}.
\end{equation}

In  \cite{Gillespie03} and  \cite{Cao06}, two successive refinements were made. The latest $\tau$-selection procedure  given in \cite{Cao06} is more accurate, easier to code, and faster to execute than the earlier procedures, but logically more complicated.

If the $\tau$ value generated above is much larger than the time required for the stochastic simulation algorithm, this approximate procedure will be faster than the exact stochastic simulation algorithm. However, it $\tau$ turns out to be less than a few multiples of the time required for the stochastic simulation algorithm to make an exact time step (in an order of $1/a_0(\vec{x})$), it would be better to use the stochastic simulation algorithm instead.

To avoid the negative populations in tau-leaping, several strategies have been proposed, in which the unbounded Poisson random numbers $k_j$ are replaced by bonded binormal random numbers  \cite{Chatterjee05, Tian04}. In 2005, Cao et al.  \cite{Cao05} proposed a new approach to resolve this difficulty.  In this new Poisson tau-leaping procedure, the reaction channels are separated into two classes: \textit{critical reactions} that may exhaust one of its reactants after some firings, and \textit{noncritical reactions} other wise. Next, the noncritical reactions are handled by the regular tau-leaping method to obtain a leap time $\tau'$. And apply the exact stochastic simulation algorithm to the critical reactions, which  gives the time $\tau''$ and the index $j_c$ of the next critical reaction. The actual time step $\tau$ is then taken to be the smaller of $\tau'$ and $\tau''$. If the former, no critical reactions fires, and if the latter, only one critical reaction $R_{j_c}$ fires. 

Below are main steps for the modified tau-leaping procedure that can avoid negative populations (refer  \cite{Cao05} for details)
\begin{enumerate}
 \item \textbf{Initialization}: Let $\vec{X}=\vec{x}_0$, and $t=t_0$.
\item \textbf{Identify the critical reactions:} Identify the reaction channels $R_j$ for which $a_j(\vec{X}) > 0$ and may exhaust one of its reactants after some firings.
\item \textbf{Calculate the leap time}: Compute the largest leap time $\tau'$ for the noncritical reactions.
\item \textbf{Monte Carlo step}: Generate $(\tau'', j_c)$ for the next critical reaction according to the modified density function according to \eqref{eq:ga}.
\item \textbf{Determine next step}: \begin{enumerate}
\item If $\tau' < \tau''$: Take $\tau = \tau'$. For all critical reactions $R_j$, set $k_j=0$. For all the noncritical reactions $R_j$, generate $k_j$ as a Poisson random variable with mean $a_j(\vec{X})\tau$.
\item If $\tau''\leq \tau'$: Take $\tau=\tau''$. Set $k_{j_c} = 1$, and for all other critical reactions set $k_{j} = 0$. For all the noncritical reactions $R_j$, generate $k_j$ as a Poisson random variable with mean $a_j(\vec{X})\tau$.
\end{enumerate}
\item \textbf{Update}: Increase the time by $\tau$, and replace the molecule count by $\vec{X} + \sum_{j=1}^M k_j\vec{v}_j$.
\item \textbf{Iterate}: Go back to Step 2 unless the simulation time has been exceeded.
\end{enumerate}

\subsection{Other simulation methods}

In additional to the prominent approximate acceleration procedure tau-leaping, there are some other strategies that tend to speedup the stochastic simulation algorithm. Here we briefly outline two of the most promising methods.

Many real systems in biological processes involve chemical reactions with different time scales, ``fast'' reactions fire very much more frequently than ``slow''ones. Procedures to handle such systems often involve a stochastic generalization of the quasi-steady-state assumption or partial (rapid) equilibrium methods
of deterministic chemical kinetics  \cite{Cao05-2, Cao05-1, Goutsias05, Rao03,  Samant05}. The \textit{slow-scale stochastic simulation algorithm} (ssSSA) (or \textit{multiscale stochastic simulation algorithm}) is a systematic procedure for partitioning the system into fast and slow reactions, and only simulate the slow reactions by  specially modified propensity functions  \cite{Cao05-2, Cao05-3, Cao05-1}. 

Another approach to simulate multiscale chemical reaction systems include different kinds of \textit{Hybrid methods} \cite{Alfonsi04, Erban06, Haseltine02, Salis05}. Hybrid methods combine the deterministic reaction rate equation with the stochastic simulation algorithm. The idea is to split the system into two regimes: the continuous regime of large molecule population species, and the discrete regime of small molecule population species. The continuous regimes is treated by ordinary differential equations, while the discrete regime is simulated by the stochastic simulation algorithm. Hybrid methods efficiently utilize the multiscale properties of the system. However, because of lacking a rigorous theoretical foundation, there are still many unsolved problems  \cite{Li08}.

There are many approaches trying to find the numerical solution of the chemical master equation  \cite{Deuflhard08, Engblom09, Engblom09-1, Hegland08, Jahnke, Jahnke08, Khoo08,  Munsky07}. In addition, numerical methods for the Langevin equation have been well documented (refer  \cite{Kloeden92} for example). We will not get into these two subjects here.

\section{Summary}

In modeling a well-stirred chemical reacting systems, the chemical master equation provides an `exact' description of the time evolution of the states. However, it is difficulty to directly study the chemical master equation because of the dimension problem. Several approximations are therefore developed, including the Fokker-Plank equation, reaction rate equation, and chemical Langevin equation. The reaction rate equation is widely used when fluctuations are not important. When the fluctuations are significant, the chemical Langevin equation can provide reasonable description for the statistical properties of the kinetics, despite the conditions in deriving the chemical Langevin equation may not hold. 

When there are noise perturbations to the kinetic parameters, there is no simple way to model the system dynamics because the time dependence of environment variables can be very complicated. In a particular case, we can replace the reaction rates by log-normal random variables, and generalize the reaction rate equation or the chemical Langevin equation to describe the dynamics of a chemical system with extrinsic noise. 

Stochastic simulation algorithm (SSA) is an `exact' numerical simulation that shares the same fundamental basis as the chemical master equation. The approximate explicit tau-leaping produce, on the other hand, is closely relate to the chemical Langevin equation. The robustness and efficiency of the two methods have been considerable improved in recent years, and these procedures seems to be nearing maturity  \cite{Gillespie07}. In the last few years, some other strategies have been developed for simulating the systems that are dynamically stiff  \cite{Gillespie07, Li08}.

\begin{figure}[htbp]
 \centering
\includegraphics[width=12cm]{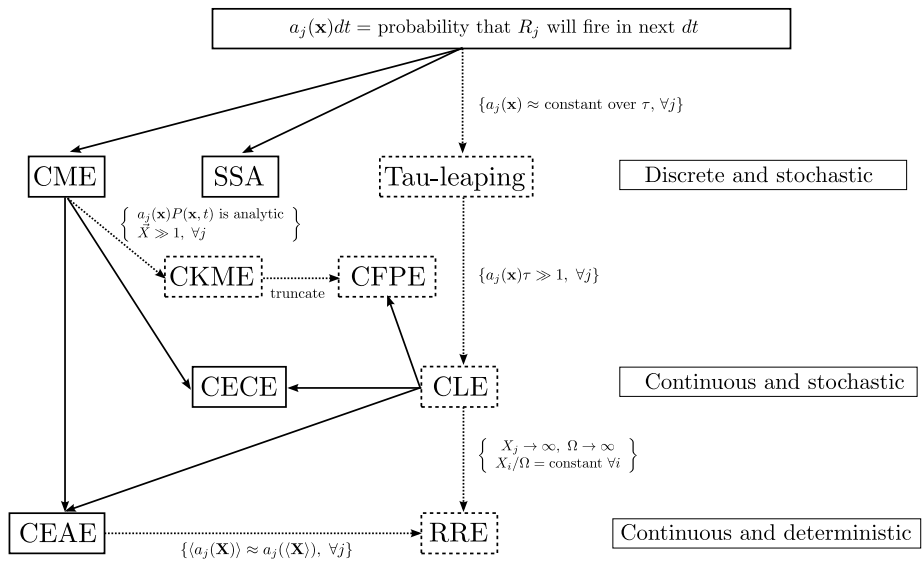}
\caption{(Modified from ref.  \cite{Gillespie07}) Theoretical structure of stochastic chemical kinetics. Everything originate from the fundamental premise of propensity function given at the top box. Solid arrows show the exact inferences routes, and dotted arrows show the approximate ones, with justified conditions in braces. Solid boxes are exact results: the chemical master equation (CME), the stochastic simulation algorithm (SSA), the chemical ensemble coefficient equation (CECE), and the chemical ensemble average equation (CEAE). Dashed boxes are approximate results: the Tau-leaping algorithm, the chemical Langevin equation (CLE), the chemical Kramers-Moyal equation (CKME), the chemical Fokker-Plank equation (CFPE), and the reaction rate equation (RRE). }
\label{fig:str}
\end{figure}

We conclude this paper with Figure \ref{fig:str}, which summarizes the theoretical structure of stochastic modeling for chemical kinetics (also refer  \cite{Gillespie07}).

\bibliographystyle{plain}
\bibliography{smsb}

\end{document}